\documentclass[12pt]{iopart}

\usepackage{mdframed}
\usepackage[normalem]{ulem} 
\usepackage{comment}
\usepackage{xcolor}
\usepackage[utf8x]{inputenc}
\usepackage[english]{babel}
\usepackage[T1]{fontenc}
\usepackage{lmodern}
\usepackage{iopams}
\usepackage[pdftex]{graphicx}
\usepackage{epstopdf}
\usepackage{braket}
\usepackage[bottom]{footmisc}
\usepackage{bbold}
\usepackage{dcolumn}
\usepackage{bm}
\usepackage{color}
\usepackage{relsize,dsfont,mathrsfs,verbatim,upgreek}
\usepackage[caption = false]{subfig}

\newcommand{\Lt}{\mathcal{L}}

\newcommand{\text}[1]{\mathrm{#1}}
\providecommand{\eqref}[1]{(\ref{#1})}

\begin{document}


\title{Local energy assignment for two interacting quantum thermal reservoirs}

\author{Alessandra Colla, Bassano Vacchini and Andrea Smirne}

\address{Dipartimento di Fisica ``Aldo Pontremoli'', Universit\`a degli Studi di Milano, Via Celoria 16, I-20133 Milan, Italy}
\address{INFN, Sezione di Milano, Via Celoria 16, I-20133 Milan, Italy}

\ead{alessandra.colla@unimi.it}
\ead{andrea.smirne@unimi.it}

\begin{abstract}
Understanding how to assign internal energy, heat, and work in quantum systems beyond weak coupling remains a central problem in quantum thermodynamics, particularly as the difference between competing definitions becomes increasingly relevant. We identify two
common sets of definitions for first-law quantities that are used to describe the thermodynamics of quantum systems coupled to thermal environments. Both are conceptually non-symmetric, treating one part of the bipartition (the “system”) differently from the other (the “bath”). We analyze these in a setting where such roles are not easily assigned — two large (but finite) sets of thermal harmonic oscillators interacting with each other. We further compare them with a third set of definitions based on a local, conceptually symmetric open-system approach (“minimal dissipation”) and discuss their quantitative and structural differences. In particular, we observe that all three sets of definitions differ substantially even when the two subsystems are weakly coupled and far detuned, and that the minimal dissipation approach features distinct work peaks that increase with the coupling strength.
\end{abstract}

\section{Introduction}

Quantum thermodynamics has matured into a vital field for understanding energy exchange 
and irreversible behavior in systems dominated by quantum effects \cite{Adesso2018,Deffner2019,Campbell2025}. The field naturally shifts focus away from studying isolated quantum systems, as the core mechanisms of energy exchange and thermalization arise from coupling the system of interest to external elements such as drives or baths. As a consequence, the theory of open quantum systems \cite{Breuer2007,Vacchini2024} is often the preferred tool for the investigation of nonequilibrium thermodynamic properties of quantum systems. 

In most used frameworks \cite{Alicki1979, Kosloff2013,Talkner2020,Potts2024}, work contributions to the system's change in energy are introduced via a parameter-dependent system Hamiltonian, where the parameters are tuned during the evolution, while heat contributions are modeled by putting the system in contact with a memoryless thermal environment. This description is purely effective 
and based on the assumption that the system evolves according to the prescribed Hamiltonian and dissipation effects. 
However, it is not always clear how to identify univocally the corresponding microscopic descriptions of the system, its environment and their coupling -- for instance,
a driving on the system can be due to both a time-dependent free system Hamiltonian 
and to a structured large bosonic thermal bath \cite{Colla2024env}.
These conceptual problems are even more pronounced at stronger interaction strength, where the effective description above will not provide a faithful characterization of the dynamics in the first place. 

Although many studies have focused on the appropriate identification of irreversibility signature and entropy production in strongly-coupled open systems \cite{Alipour2016,Alipour2017,Popovic2018,Strasberg2019,Rivas2020,Colla2022a}, the ambiguities in defining thermodynamic quantities starting from a microscopic description of the system and environment begin already at the level of the first law. The definitions of heat exchange, work and even internal energy are inherently ambiguous when the interaction energy is non-negligible and can be accounted for in different ways. 
This issue has remained relatively underground in mainstream literature, with some works addressing it directly by developing different approaches to the first law \cite{Weimer2008,Alipour2016,Alipour2022,Colla2022a,Elouard2023}; it has now, however, become more widely recognized, and comparisons between different definitions of first law quantities are being carried out \cite{Seegebrecht2024,Davoudi2025,Picatoste2025}. 

In the literature, one can find two different sets of definitions which both rely on the identification of heat as the change in energy of the bath \cite{Esposito2010,Landi2021}, but differ in the treatment of internal energy and work. Both approaches are widely used and rely on the asymmetric labeling of each side of the bipartition: namely, one needs to define in advance which subsystem is the ``system'' and which is the ``bath''. While never explicitly stated, it seems that an acceptable bath is typically modeled via a large set of bosonic or fermionic modes initialized in a thermal state -- though the assignment of the ``bath'' role has also been given to small quantum systems, such as a single qubit in the context of collisional models \cite{Rodrigues2019}.

In this work, we compare these two sets of definitions by applying them to both sides of a bipartition given by two quantum systems that could in principle be considered ``baths'': namely, two large sets of interacting bosonic modes, each initialized in a thermal state at different temperatures. 
To the comparison, we add a third set of definitions for first law quantities given by a recent approach that relies on open quantum system techniques and is applicable to general environments (i.e., the other side of the bipartition need not necessarily be a ``bath'') \cite{Colla2022a}. As such, the approach can be equally well applied to each set of modes, and indirectly addresses the problem of where to place the interaction energy by predicting a renormalization of the energy levels of each subsystem due to the interaction. None of the definitions considered are expected to satisfy some kind of energy balance at strong coupling; i.e., it is not expected that heat, work and internal energies computed this way flow from one system to the other as in classical thermodynamics. We nonetheless gauge this situation to see whether such a balance holds (even approximately) in the case of two ``baths''.

The work is structured as follows: in Sec.~\ref{sec:approaches}, we present the three different sets of definitions for internal energy, work and heat. In Sec.~\ref{sec:model} we provide details of the model considered and how to apply the three approaches in this case, while in Sec.~\ref{sec:analytical} we select a specific regime of parameters where a fully analytical treatment is possible, in order to define quantities and regimes of interest. We show our results, including comparison and specific features of the different approaches, in Sec.~\ref{sec:results}.
Finally, concluding remarks and future perspectives are given in Sec.~\ref{sec:con}.

\section{Approaches to the first law}\label{sec:approaches}

Let the total system (i.e., the system-environment pair) be described by a Hamiltonian of the following form:
\begin{equation}
H = H_S(t) + H_E + H_I \; .
\end{equation}
We assume the interaction Hamiltonian $H_I$ to be time independent (no modulation or switch on/off) to better understand how the assignment and placement of the variation of interaction energy affects the definitions of the thermodynamic quantities. In our model of interest, the system Hamiltonian will also be taken time-independent ($H_S(t)\equiv H_S$), so that there are no external driving protocols acting on either side of the bipartition, and the global system becomes fully autonomous.
As a consequence of this choice, some of the definitions will predict a zero work contribution. However, the frameworks described here are applicable also to systems under driving or with modulated interaction, which are typically responsible for further work-like contributions. 

To introduce the first two sets of definitions, it is useful to start from the following ``internal energy-like'', ``work-like'' and ``heat-like'' rates associated to each term of the global Hamiltonian:
\begin{eqnarray}
\dot{U}_X(t) & = & \frac{d}{dt}\Tr \{{H}_X \rho_{SE}(t)\} \; , \\
\dot{Q}_X(t) & = & \Tr \{{H}_X \dot{\rho}_{SE}(t)\} \; ,  \\
\dot{W}_X(t) & = & \Tr \{\dot{{H}}_X \rho_{SE}(t)\} \; , 
\end{eqnarray}
where $X=S,E,I$ and $\rho_{SE}(t)$ is the global state evolved at time $t$. These contributions are formal and not really associated to physical quantities, but by definition each set labeled by $X$ satisfies a first-law-like relation, namely $\dot{U}_X(t) = \dot{Q}_X(t)+ \dot{W}_X(t)$. Moreover, it holds that the global heat contribution is zero because the total system+environment compound is a closed quantum system: $\dot{Q}_S(t)+\dot{Q}_E(t)+\dot{Q}_I(t)=0$. 
The traditional first law of quantum thermodynamics for weakly-coupled open quantum systems \cite{Alicki1979, Kosloff2013} focuses on the terms with $X=S$:
in the case we will study, where no work protocol is imposed on the system, they read
\begin{eqnarray}
    \Delta U^{\text{wc}}_S(t) &=&  U_S(t) - U_S(0) \; ,\\
    \delta Q^{\text{wc}}_S(t)  &=& U_S(t) - U_S(0) \equiv \Delta U^{\text{wc}}_S(t)\\
     \delta W^{\text{wc}}_S(t)  &=& 0 \; .
\end{eqnarray}
In this regime, the interaction energy is supposed to be negligible, so that, in the absence of any driving involving the environment, 
$\dot{Q}_E(t) = -\dot{Q}_S(t)$
and the energy balance for the open system is perfectly mirrored by the balance for the environment
via a simple change of sign of the corresponding thermodynamic quantities. Indeed, beyond the weak coupling regime, when the interaction energy
is no longer negligible, such a simple picture has to be abandoned.

The first approach we consider to deal with the strong-coupling regime is the one found originally in Ref.~\cite{Esposito2010}, where the heat exchanged with the system is identified with the change of energy leaving the environment (i.e., taken with a minus sign). Its rate is therefore given by $-\dot{U}_E(t)$. This choice allows one to write an entropy production which is always positive for systems that start uncorrelated with a ``bath'' (any non-driven quantum system initialized in a thermal state). Because of the underlying assumption that there is no work protocol being performed on the bath, it holds that $-\dot{U}_E(t)= -\dot{Q}_E(t)= \dot{Q}_S(t)+\dot{Q}_I(t)$. Therefore, the heat-like term $\dot{Q}_I(t)$ due to the interaction Hamiltonian is taken to contribute to the heat entering the system. Moreover, since the driving protocol is the only thing responsible for the change of energy of the global (closed) system, it must lead to work-like contribution; because the driving is applied only to the system degrees of freedom, it is fully associated with the system whose work contribution is given by $\dot{U}_S(t) + \dot{U}_E(t)+ \dot{U}_I(t) = \dot{W}_S(t)$. Then, to make sure that the first law of quantum thermodynamics holds for the system of interest, one is forced to define the internal energy of the system as also including the interaction energy variation, namely $\dot{U}_S(t)+\dot{U}_I(t)$.

This approach to first law quantities is the only one in this study which is compatible with the notion that work done on the global system is equivalent to the work done on the open system (for time-independent interaction), which is a necessary condition for certain derivations of the Jarzynski equality at strong coupling \cite{Talkner2020}. In the case where the system of interest has no external work protocol performed on it, as in the case we will consider, this approach -- which we label \emph{the interaction approach} for convenience -- gives the following thermodynamic quantities (energy changes from time 0 to time $t$):
\begin{eqnarray}
\Delta U^{\text{int}}_S(t) &=&  U_S(t) - U_S(0) + U_I(t) - U_I(0) = - U_E(t) + U_E(0)\; ,\label{eq:intu}\\
\delta Q^{\text{int}}_S(t)  &=& - U_E(t) + U_E(0) \equiv  \Delta U^{\text{int}}_S(t) \; ,\\
\delta W^{\text{int}}_S(t)  &\equiv &  0\; .
\end{eqnarray}

The second approach we consider \cite{Landi2021,Popovic2021,Popovic2023} hinges on the same definition of heat as the interaction approach, but does not include the interaction contribution in the internal energy of the system, which is simply taken to be $\dot{U}_S(t)$; for this reason, we label it as \emph{the bare approach}. 
The work contribution is defined by imposing the first law, so that now it includes the contribution $\dot{U}_I(t)$. In our case of interest, with no external work done on the system, this leads to
\begin{eqnarray}
    \Delta U^{\text{bare}}_S(t) &=&  U_S(t) - U_S(0) \; ,\\
    \delta Q^{\text{bare}}_S(t)  &=& - U_E(t) + U_E(0)  \equiv   \delta Q^{\text{int}}_S(t),\\
     \delta W^{\text{bare}}_S(t)  &=& - U_I(t) + U_I(0) \equiv \; \Delta U^{\text{bare}}_S(t) - \delta Q^{\text{bare}}_S(t). \label{eq:barew}
\end{eqnarray}
Notice that only the variation of energies $U_E$ and $U_S$ are needed to compute all the thermodynamic quantities for both the interaction and the bare approach in Eqs.(\ref{eq:intu})-(\ref{eq:barew}). In general, since the environment is often an infinitely large quantum system, its energy variation $U_E$ can be hard to access, requiring advanced techniques or finite-size approximations \cite{Pucci2013,Kato2016,Colla2021,Gribben2022,Albarelli2024}.

The third and last approach we consider in this study is the one proposed in \cite{Colla2022a} and is based on open quantum system techniques. The thermodynamic quantities are all defined based on the reduced evolution of the system and the information found within; 
no evaluation of quantities like $U_E(t)$ or $U_I(t)$ are necessary. In particular, it is not necessary to access any part of the state evolution except for the reduced one. The approach assumes that one is able to control the system of interest well enough such that one has access to the map propagating the reduced system state in time $\Phi_t[\rho_S(0)]=\rho_S(t)$. Given this evolution, its time-local generator -- given by $\Lt_t = \dot{\Phi}_t \Phi_t^{-1}$ -- describes a general open system evolution $\dot{\rho}_S(t) = \Lt_t[\rho_S(t)]$ (also in presence of strong coupling and memory effects), and can be written in the following generalized Lindblad form \cite{Breuer2012,Hall2014}:
\begin{equation}
\hspace*{-\mathindent}\Lt_t[\rho_S(t)] = -i [ K_S(t), \rho_S(t)] + \sum_k \gamma_k(t) \left[ L_k(t)\rho_S(t)L_k^\dag(t) -\frac{1}{2} \left\{L_k^\dag(t) L_k(t), \rho_S(t) \right\} \right] \; .
\end{equation}
This equation in general features 
a time-dependent Hamiltonian operator $K_S(t)=K^{\dagger}_S(t)$ responsible for the unitary part of the dynamics, as well as a dissipator with time-dependent Lindblad operators $L_k(t)$ and rates $\gamma_k(t)$, which can in principle be also negative and account for non-Markovian effects. 

The separation of the generator into a Hamiltonian and a dissipative part is highly non-unique \cite{chruscinski2022}, 
but asking for an additional minimization condition on the dissipator as a superoperator \cite{Sorce2022} -- also termed as \emph{principle of minimal dissipation} in \cite{Colla2022a} -- singles out one specific choice. 
The resulting system Hamiltonian $K_S(t)$, which acts as a renormalized energy operator for the system, has been used in \cite{Colla2022a}
to define the thermodynamic properties of the system. Because the effective Hamiltonian $K_S(t)$ can in principle be time-dependent even in a fully autonomous global system (no external work protocol on $S$ \cite{Colla2025a}), the general definitions for the first-law quantities, which are here grouped under the name \emph{minimal dissipation approach}, read 
\begin{eqnarray}
    \Delta U^{\text{md}}_S(t) &=&  \Tr\{{K}_S(t)\rho_S(t)\} - \Tr\{{K}_S(0)\rho_S(0)\} \; ,\\
    \delta Q^{\text{md}}_S(t)  &=& \int_0^t d\tau \Tr\{{K}_S(\tau)\dot{\rho}_S(\tau)\},\\
     \delta W^{\text{md}}_S(t)  &=& \int_0^t d\tau \Tr\{\dot{{K}}_S(\tau)\rho_S(\tau)\} \; .
\end{eqnarray}
These definitions can be seen as a straightforward extension to the effective, weak-coupling description, as they lead precisely to those definitions as $K_S(t)$ approaches $H_S(t)$ in the weak-coupling and Markovian limit \cite{Albash2012}.

All above sets of quantities refer to the system of interest $S$, while all other degrees of freedom are considered ``the environment''. As we already mentioned, there might be situations where there is no natural bipartition where one side can be clearly called ``the system'' and the other ``the environment''. Particularly in our selected case of interest, where two large sets of bosonic modes start in a thermal state, both sides of the bipartition could act as the environment. We therefore consider each subset of oscillators as ``the system'' and the other as ``the environment'', and swap them to provide a thermodynamic treatment for both sides of the bipartition.

\section{Bipartite model for thermal exchange}\label{sec:model}
In this section, we introduce our bipartite model of two large sets of bosonic modes, and work through the steps of solving the global dynamics as well as recovering the reduced system master equation (for each subset), in order to be able to compute exactly all three sets of definitions for work, heat and internal energy. 
We stress that, compared to previous studies \cite{Marinez2013,Esposito2015,Nicacio2016,Leitch2022,Razzoli2024} on thermodynamics for linear bipartite quantum systems, 
we are here investigating the simultaneous evolution of thermodynamics quantities of both parts, and not focusing on the impact of one of the two (``the environment'')
on the other (``the system'').

\subsection{Exact dynamics}
\begin{figure}[tp]
\centering
\includegraphics[width=\textwidth]{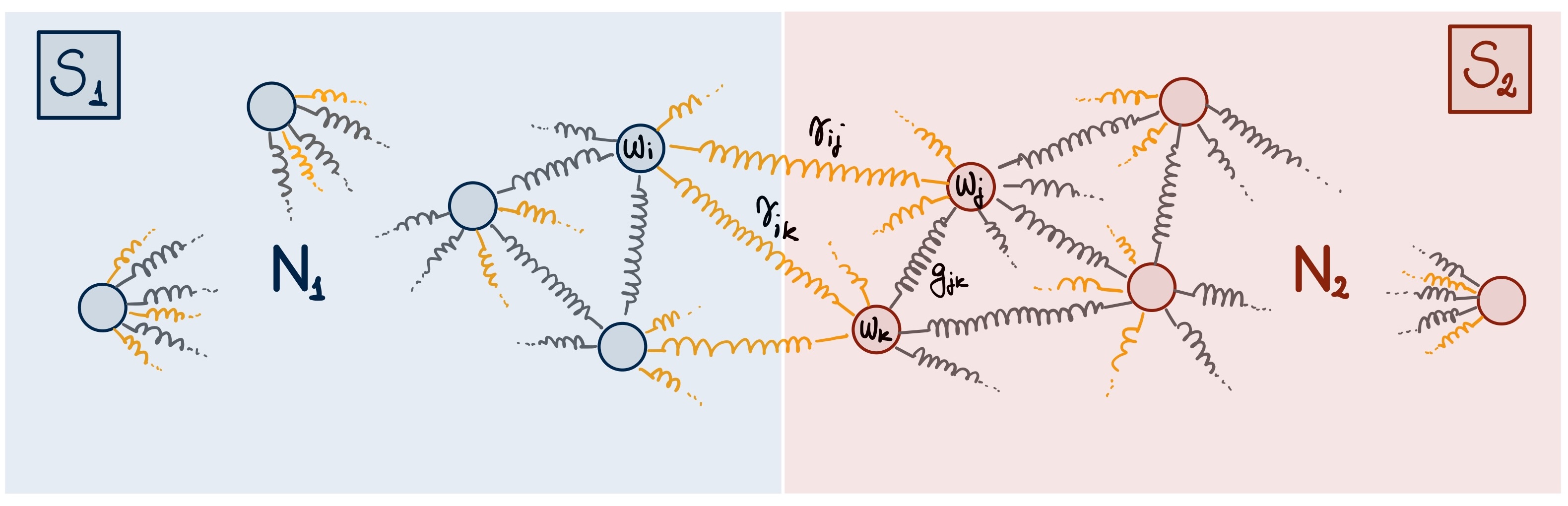}
\caption{Sketch of the model considered in this work: two sets of bosonic modes, all linearly coupled as described by the Hamiltonian of Eq.~\eqref{eq:HNM}.}
\label{fig:sketch}
\end{figure}

We consider two coupled sets of bosonic modes (or equivalently harmonic oscillators): one (system $S_1$) made up of $N_1$ oscillators, and the other (system $S_2$) of $N_2$. Each oscillator in $S_{1(2)}$ interacts with each oscillator in $S_{2(1)}$ through exchange linear coupling, and all oscillators within $S_{1(2)}$ interact with each other. The global Hamiltonian reads
\begin{eqnarray}\label{eq:HNM}
    \hat{H} = \underbrace{\sum_{ij=1}^{N_1} h^{(1)}_{ij} a^\dag_i a_j}_{\hat{H}_1} + \underbrace{\sum_{kl=N_1+1}^{N_1+N_2}  h^{(2)}_{kl} a^\dag_k a_l}_{\hat{H}_2} + \underbrace{\sum_{i=1}^{N_1} \sum_{k=N_1+1}^{N_1+N_2} \gamma_{ik} ( a^\dag_i a_k +a^\dag_k a_i)}_{\hat{H}_I} \; ,
\end{eqnarray}
where $a_j$ and $a_j^\dag$ are annihilation and creation operators of the mode $j$; see Fig.~\ref{fig:sketch} for a sketch of the model.

Since both systems start in thermal states and the Hamiltonian is quadratic, the total system starts and remains in a Gaussian state. Its dynamics can be easily solved via exact diagonalization of the Hamiltonian matrix (i.e. the transformation to normal modes). To do so, we define a vector of $(N_1+N_2)$ operators
\begin{eqnarray}\label{eq:vecR}
    \vec{R} := (a_1, a_2, ... , a_{N_1}, a_{N_1+1}, ... , a_{N_1+N_2})^T\; 
\end{eqnarray}
and a Hermitian -- symmetric if all couplings are taken real -- $(N_1+N_2)\times(N_1+N_2)$ Hamiltonian matrix
\begin{eqnarray}    
   H =\left[\begin{array}{cc}
H_1 & G \\
G^T & H_2 
\end{array}\right] \; ,
\end{eqnarray}
where $H_{1} = \{h^{(1)}_{ij}\}_{i,j =1,...,N_1}$ is a $N_1\times N_1$ matrix, analogously $H_{2} = \{h^{(2)}_{kl}\}_{k,l =N_1+1,...,N_1+N_2}$ is a $N_2\times N_2$ matrix, and $G = \{\gamma_{ik}\}_{i=1,...N_1, k= N_1+1,...,N_1+N_2}$ is the $N_1\times N_2$ inter-coupling matrix. We can thus rewrite the Hamiltonian operator as
\begin{eqnarray}
    \hat{H} = (\vec{R}^\dag)^T {H} \vec{R} \; ,
\end{eqnarray}
where the dagger is understood as acting on the operators, and the transpose only on the vectorial structure.

The matrix ${H}$ can be diagonalized by an orthogonal matrix $Z$ 
\begin{eqnarray}
    H' = Z HZ^T = \text{diag}\{z_1, z_2, ... , z_{N_1}, z_{N_1+1}, ... , z_{N_1+N_2}\}, \; 
\end{eqnarray}
which transforms the system into normal modes, where the Hamiltonian operator now reads
\begin{eqnarray}
    \hat{H} = (\vec{R'}^\dag)^T {H'} \vec{R'} \; 
\end{eqnarray}
with transformed vectors $ \vec{R'} = Z \vec{R}$.
Notice that this transformation leaves the commutation relations of the modes invariant. 
Using this diagonalization we can solve the transformed system using Heinsenberg's equations:
\begin{eqnarray}
    \frac{d}{dt} a'_j (t) = -i z_j a'_j(t) \; , \quad \forall \; j = 1, ..., N_1+N_2 \; , 
\end{eqnarray}
and thus
\begin{eqnarray}
    a'_j (t) = e^{-i z_j t} a'_j(0) \; , \quad \forall \; j = 1, ..., N_1+N_2 \; .
\end{eqnarray}
Writing this in matrix form as $\vec{R'} (t) = e^{-i H' t}\vec{R'}(0)$ and transforming back and forth with $Z$, we obtain the solution for the original Heisenberg operators as well as their time derivative:
\begin{eqnarray}
    \vec{R} (t) = Z^T e^{-i H' t}Z\vec{R}(0) \;, \quad \quad \frac{d}{dt} \vec{R} (t) = -i Z^T H' e^{-i H' t}Z\vec{R}(0) \;.
\end{eqnarray}

Using the above strategy, one can easily compute any first or second moment of interest (including correlations between the two subsystems). Since the global state is Gaussian, they are the only quantities needed to determine the exact dynamics. The first moments (a $(N_1+N_2)$-dimensional vector) are given in terms of the initial first moments by
\begin{eqnarray}
    \braket{\vec{R}}_t = Z^T e^{-i H' t}Z \braket{\vec{R}}_0 \;,
\end{eqnarray}
while the second moments are
\begin{eqnarray}\label{eq:S-evolved}
    \braket{\vec{R}^\dag(\vec{R})^T}_t &= Z^T e^{i H' t}Z  \braket{\vec{R}^\dag(\vec{R})^T}_0 Z^T e^{-i H' t}Z \\
    \braket{\vec{R}(\vec{R})^T}_t &= Z^T e^{-i H' t}Z  \braket{\vec{R}(\vec{R})^T}_0 Z^T e^{-i H' t}Z  \; ,
\end{eqnarray}
where the above equations indicate the $[(N_1+N_2)\times(N_1+N_2)]$ matrices with entries $\braket{a^\dag_i a_j}_t$ and $\braket{a_i a_j}_t$, respectively.

\subsection{The reduced dynamics}

Using the exact dynamics evaluated for the global system, we can find the reduced dynamics -- and the exact master equation generating it -- for each of the subsystems by tracing out the other. We show here the procedure for subsystem $S_1$, but it applies identically for subsystem $S_2$. Let us use the notation
\begin{eqnarray}\label{eq:vecr12}
    \vec{r}_1 := (a_1, a_2, ... , a_{N_1})^T\; \quad \quad    \vec{r}_2 := (a_{N_1+1}, ... , a_{N_1+N_2})^T\; 
\end{eqnarray}
so that $\vec{R} = (\vec{r}_1^T, \vec{r}_2^T)^T $. Then, the operation of partial trace of $S_2$ consists in simply removing the last $N_2$ rows of the vector: $\vec{r}_1 = \Tr_2 \{ \vec{R}\}$. 
  
To define the evolution operator for $S_1$, we first need to establish the initial conditions for the ``environment'', namely the other subsystem $S_2$. 
We fix the initial state for $S_2$ as a thermal state, which consists in having first moments set to zero 
\begin{eqnarray}\label{eq:env-firstm-th}
    \braket{\vec{r}_2}_0 = \braket{\vec{r}_2(\vec{r}_2)^T}_0= 0, \;
\end{eqnarray}
while the second moments matrix $\braket{\vec{r}_2^{\dag}(\vec{r}_2)^T}_0$ can be found again with a back and forth transformation to normal modes (see \ref{sec:appA}).

We can then write down the evolution of the first moments of system $S_1$ as a function of their initial values, and find their evolution matrix $\Phi_t$ by extending the operation of partial trace to matrices (namely removing all last $N_2$ rows and columns):
\begin{eqnarray}
    \braket{\vec{r}_1}_t &=& \Phi_t \braket{\vec{r}_1}_0  = \Tr_2\{ Z^T e^{-i H' t}Z \braket{\vec{r}_1}_0 \oplus \braket{\vec{r}_2}_0\} \nonumber\\
    &=& \Tr_2\{ Z^T e^{-i H' t}Z \} \braket{\vec{r}_1}_0 \;,
\end{eqnarray}
where $\oplus$ denotes the direct sum and the last step holds because of \eqref{eq:env-firstm-th}. From this, it follows that
\begin{eqnarray}\label{eq:firstm-d}
    \frac{d}{dt}  \braket{\vec{r}_1}_t = \dot\Phi_t  \braket{\vec{r}_1}_0 = -i \Tr_2\{Z^T H' e^{-i H' t}Z\} \braket{\vec{r}_1}_0\;.
\end{eqnarray}

The matrix $\Phi_t$ is actually sufficient to find the effective Hamiltonian for the reduced system. 
Indeed, for the microscopic model considered, the master equation is of the following form \cite{Xiao2013}:
\begin{eqnarray} \nonumber
    \Lt_t [X] &=&  -i [ \sum_{ij=1}^N K_{ij}(t)a_i^\dag a_j , X]  + \sum_{ij=1}^N \Gamma_{ij}^+(t) [ a_i^\dag X a_j -\frac{1}{2} \{a_j a_i^\dag ,X \}] +\\ & & + \sum_{ij=1}^N \Gamma_{ij}^-(t) [ a_j X a_i^\dag -\frac{1}{2} \{a_i^\dag a_j,X \}]. \label{eq:tcl-model}
\end{eqnarray}
This is a consequence of the fact that the environment is initially thermal (a zero mean Gaussian state) and that the total evolution brings Gaussian states into Gaussian states. 
Deriving the evolution equation of first moments from the master equation above,
\begin{eqnarray}
    \frac{d}{dt}  \braket{\vec{r}_1}_t = L_t\braket{\vec{r}_1}_t :=  \left[-i K_t + \frac{1}{2} \Gamma^+_t -  \frac{1}{2} \Gamma^-_t \right]\braket{\vec{r}_1}_t\;,
\end{eqnarray}
we find that the effective Hamiltonian matrix $K_t = \{K_{ij}(t)\}_{ij}$ appears as the imaginary part of the generator matrix $L_t$ for the first moments.

By making explicit the evolution matrix $\Phi_t$ in the above we have that $\frac{d}{dt}  \braket{\vec{r}_1}_t =  L_t \Phi_t \braket{\vec{r}_1}_0$, which, when compared with eq.~\eqref{eq:firstm-d}, gives
\begin{eqnarray}\label{eq:Mt-firts}
     L_t = \dot{\Phi}_t\Phi_t^{-1} = -i \Tr_2\{Z^T H' e^{-i H' t}Z\} \times \left[ \Tr_2\{ Z^T e^{-i H' t}Z \}\right]^{-1}  \;.
\end{eqnarray}
The imaginary part of the above therefore gives the effective Hamiltonian matrix:
\begin{eqnarray}\label{eq:Kt}
    K_t = \frac{L^\dag_t-L_t}{2 i} \; ,
\end{eqnarray}
which is completely determined by the first-moment evolution. The rates matrices $\Gamma^+_t$ and $\Gamma^-_t$ are harder to determine, as they require the comparison with the evolution of second moments using the same method as above. We however do not derive them explicitly here, 
as they are not needed for the thermodynamic quantities of interest.
    
\subsection{Thermodynamic quantities}
The dynamics of both the global system and the two reduced subsystems determine the thermodynamic quantities we are interested in. Let us start with the bare energies of each subsystem (namely, the expectation value of their bare Hamiltonian). Defining the notation $S_t = \braket{\vec{R}^\dag (\vec{R})^T}_t$, the expectation value of $\hat{H}_1$ and $\hat{H}_2$ at any point in time are given by
\begin{eqnarray}
    E_x(t) = \Tr\{ H_x \Tr_{\bar{x}} \{S_t \} \} \; ,
\end{eqnarray}
where, here and in the remainder of the work, we use the labels $x=1,2$ and $\bar{x}=2,1$, respectively.

From these energies, the first-law quantities given by the weak coupling formulation are trivial.
They can also be used directly to find two of the different sets of definitions considered for strong coupling in Sec.~\ref{sec:approaches}. Internal energy, work and heat according to the interaction approach are given by
\begin{eqnarray}
    \Delta U^{\text{int}}_x(t) &=& \delta Q^{\text{int}}_x(t) =
    -E_{\bar{x}}(t)+E_{\bar{x}}(0)
    =  \Tr\left\{ H_{\bar{x}} \Tr_x \left\{S_0  -S_t \right\}  \right\}  \; ,\\
     \delta W^{\text{int}}_x(t)  &\equiv& 0\; .
\end{eqnarray}
Notice how there is no work on either subsystem (independently on the coupling strength and shape) and the fact that the total energy change is not additive by definition:
\begin{eqnarray}
    \Delta U^{\text{int}}_1(t)+\Delta U^{\text{int}}_2(t) = \Delta U_I(t) \; .
\end{eqnarray}
The definitions of the bare approach instead give the following
\begin{eqnarray}
    \Delta U^{\text{bare}}_x(t) &=&  E_x(t)-E_x(0)  = \Tr\left\{ H_x \Tr_{\bar{x}} \left\{S_t  -S_0 \right\}  \right\} \; ,\\
    \delta Q^{\text{bare}}_x(t)  &=&  \delta Q^{\text{int}}_x(t) =  \Tr\left\{ H_{\bar{x}} \Tr_x \left\{S_0  - S_t \right\}  \right\} \;,\\
     \delta W^{\text{bare}}_x(t)  &=& \delta W^{\text{bare}}_{\bar{x}}(t)=\Tr\left\{ H_1\oplus H_2  \left[S_t  -S_0 \right]  \right\} \; .
\end{eqnarray}
The same note on energy additivity applies here, but notice how the final balance has opposite sign with respect to the interaction approach
\begin{eqnarray}
    \Delta U^{\text{bare}}_1(t)+\Delta U^{\text{bare}}_2(t) = -\Delta U_I(t) \; .
\end{eqnarray}
In contrast to the interaction approach, this framework allows for work being performed on each of the subsystems. However, this work is the exact same contribution for both subsystems (with the same sign), so neither heat nor work balance out:
\begin{eqnarray}\label{eq:balance-bare}
  \hspace*{-2em}&  \delta Q^{\text{bare}}_1(t)+\delta Q^{\text{bare}}_2(t) = \Delta U_I(t) \; ,
   & \;\; \delta W^{\text{bare}}_1(t)+\delta W^{\text{bare}}_2(t) = - 2 \Delta U_I(t) \; .
\end{eqnarray}

The last set of thermodynamic quantities we need is from the minimal dissipation approach. In contrast to the other definitions, they cannot be calculated using the bare energies $U_x$; they center on the effective Hamiltonian $K^{(x)}_t$ of each subsystem $x$ and involve time-derivatives and integrals:
\begin{eqnarray}
    \Delta U^{\text{md}}_x(t) &=& \Tr\left\{ K^{(x)}_t \Tr_{\bar{x}} \left\{S_t \right\} \right\} - \Tr\left\{K^{(x)}_0  \Tr_{\bar{x}} \left\{S_0 \right\}  \right\}\; ,\\
    \delta Q^{\text{md}}_x(t)  &=& \int_0^t d\tau \Tr\left\{ K^{(x)}_\tau \Tr_{\bar{x}} \left\{\frac{d}{d\tau}S_\tau \right\} \right\} \;,\\
     \delta W^{\text{md}}_x(t)  &=& \int_0^t d\tau \Tr\left\{ \frac{d}{d\tau}K^{(x)}_\tau \Tr_{\bar{x}} \left\{S_\tau \right\} \right\}  \;.
\end{eqnarray}
In general, balance relations are known to not always hold for minimal dissipation \cite{Neves2024}. From these equations alone, moreover, it is unclear if and how much they deviate from them with respect to the other definition sets. Of course, for vanishing interaction all definitions for all approaches trivially balance out; however, there might be cases where the interaction energy variation is identically zero -- which would reconcile the bare and the interaction approach, that also then balance out -- but where the minimal dissipation approach assigns energetic contributions that do not balance out \cite{Neves2024}.

\section{Homogeneous coupling and frequencies}\label{sec:analytical}
We consider now the case where all the oscillators of the first subsystem have frequency $\omega_1$ and intra-coupling $g_1$; similarly for the second subsystem using $\omega_2$ and $g_2$. We also assume the inter-coupling between the two subsystems to be the same for each pair of oscillators, namely given by a unique value $\gamma$. This case can be fully solved analytically -- see \ref{sec:appA} -- and gives us an idea of how the different parameters influence the dynamics and thermodynamics of the systems.

\subsection{Relevant model parameters and internal energies}
Looking first at the bare subsystems (before they are coupled to each other), the eigenvalues of each subsystem Hamiltonian matrix determine their bare energy. Let $x=1,2$ label each subsystem, which has bare Hamiltonian matrix $H_x$. In its spectrum, there are only two eigenvalues, one with $N_x-1$ degeneracy, and one with no degeneracy:
    \begin{eqnarray}
     \hspace*{-2em}   \epsilon_x = \omega_x - g_x \quad (\text{deg: } N_x-1) \; , \quad \quad
        \nu_x = \omega_x + (N_x-1)g_x \quad (\text{deg: } 1) \; .
    \end{eqnarray}
The non-degenerate eigenvalues $\nu_{1,2}$ turn out to be the only ones entering energy variations. 

Proceeding with the full diagonalization of the model (with coupling), we obtain a spectrum of four distinct eigenvalues. Two of them are the $N_1-1$- and $N_2-1$-degenerate eigenvalues of the bare subsystems ($\epsilon_1$ and $\epsilon_2$, respectively) and are associated to eigenvectors that live on the space of subsystems only, while the other two eigenvectors live in both subspaces (and are thus responsible for the interaction between $S_1$ and $S_2$) and have eigenvalues:
\begin{eqnarray}
        \lambda = \frac{\nu +\Omega}{2} \; , \quad \quad
        \mu = \frac{\nu -\Omega}{2} \; .
\end{eqnarray}
In the expressions above, we have defined the following quantities:
\begin{eqnarray} \label{eq:nuomega}
        \nu = {\nu_1+\nu_2} \; , \quad \quad
        \Omega = \sqrt{\Delta^2+\Gamma^2} \; ,
\end{eqnarray}
in terms of an effective detuning $\Delta$ and coupling strength $\Gamma$:
\begin{eqnarray}
        \Delta = \nu_1-\nu_2 \; ,\quad \quad
        \Gamma =2 \sqrt{N_1 N_2} \gamma \; .
\end{eqnarray}

Given these parameters and the initial inverse temperature of each subsystem ($\beta_{1,2}$), the internal-energy variations according to each approach are: 
\begin{eqnarray} \label{eq:Ubare-a}
&&   \Delta U_x^{\text{bare}}(t) = \nu_x G_x(t) \; , \\ \label{eq:Uint-a}
&&    \Delta U_x^{\text{int}}(t) = -\nu_{\bar{x}} G_{\bar{x}}(t)\; , \\ \label{eq:Umd-a}
&&   \Delta U_x^{\text{md}}(t) = \nu_x G_x(t) +  (\nu_x-\nu_{\bar{x}}) B(t) \left[ n_x(\nu_x) + G_x(t)\right]\; ,
\end{eqnarray}
where we have defined $n_x(X) = (e^{\beta_x X}-1)^{-1}$ and
\begin{eqnarray}
G_1(t)&=& -G_2(t)= (n_2(\nu_2) - n_1(\nu_1)) \frac{\Gamma^2}{\Omega^2} \sin^2\left(\frac{\Omega}{2} t\right)\; ,  \\
B(t) &=& -\frac{\Gamma^2 \sin^2\left(\frac{\Omega}{2} t\right)}{2\left[\Delta^2+\Gamma^2\cos^2\left(\frac{\Omega}{2} t\right)\right]}  \; . 
\end{eqnarray}

Notice that since $B(t)\leq0$ and $n_x(\nu_x)+G_x(t)\geq0$, the sign of the detuning $\Delta$ determines whether the internal energies of minimal dissipation are greater or smaller than the ones from the bare approach. Indeed, the effective detuning plays often a decisive role in these quantities, as we will see in Sec.~\ref{sec:results}.

The main frequency of oscillation for all quantities is the frequency $\Omega = \sqrt{\Delta^2+\Gamma^2}$ (analog to a Rabi frequency), and constitutes the only frequency of interest for the energies of the bare and interaction approaches, which are proportional to $\sin^2(\Omega t/2)$ and thus reminiscent of Rabi oscillations. The minimal dissipation energies, instead, present more complex behaviour, due to the dependency on the backward propagator. We will see in Sec.~\ref{sec:results} how this will lead to a systematic difference in the approaches at stronger coupling.

\subsection{Regimes of interest}\label{sec:regimes}
From the quantities above and the main parameters governing their evolution, it is possible to distinguish some specific parameter regimes. Different regimes are determined by the interplay between the effective detuning $\Delta$ and the coupling strength $\Gamma$. Because of the analogy to Rabi oscillations in the analytical case, we consider the two extreme regimes:

\begin{itemize}
\item \textit{Dispersive regime --} The two subsystems are far detuned with respect to the strength of the interaction:
\begin{eqnarray}
    \Delta\gg \Gamma \; .
\end{eqnarray}
It means that energy is exchanged inefficiently between the two subsystems, and we thus consider it as a weak-coupling regime. Note, however, that this is very different to the Markovian regime for open quantum systems coupled to large baths, where one expects the spectral density to be flat around the main frequency of the system. 
By looking at $B(t)$, which plays a key role in the difference between the minimal-dissipation energy and the bare energy, in this regime we have
\begin{eqnarray}
    B(t)  \stackrel{\Delta \gg \Gamma}{\approx} - \frac{\Gamma^2}{2\Delta^2}\sin^2\large(\frac{\Omega}{2} t\large) \; ;
\end{eqnarray}
we note that: (1) the secondary frequencies in the behavior of renormalized energies are suppressed, (2) the magnitude of $B(t)$ is small. Notice that, however, also the detuning is featured in the extra term in the minimal dissipation energies \eqref{eq:Umd-a}; as we will see later, this will contribute to the fact that, contrary to intuition, the different definitions of internal energy are actually all incompatible even in the dispersive regime. 

\item \textit{Ultrastrong coupling regime --} Based on the dispersive regime analysis, we expect strong coupling effects to appear when $\Gamma$ is non-negligible with respect to $\Delta$, and energy exchanges are more efficient due to higher coupling and resonance effects. Therefore, we consider the coupling to be ultrastrong whenever
\begin{eqnarray}
    \Delta \ll \Gamma \; .
\end{eqnarray}
In this regime, there is a starker difference between the minimal dissipation energy and the bare energy. Indeed, for times that are not too close to $t=(2n+1)\pi/(\Omega)$
\begin{eqnarray}
    B(t) \stackrel{\Delta \ll \Gamma}{\approx} - \frac{1}{2}\tan^2(\Omega t) \; ,
\end{eqnarray}
which is responsible for additional peaks in the minimal dissipation quantities. These peaks are smoothed out around divergencies, the sooner the farther one is from the ultrastrong limit, and have a direction determined by the sign of $\Delta$.
\end{itemize}

A different set of regimes, instead, deals with whether the subsystems act as a collective mode or not. It has nothing to do with the coupling between the two sets of oscillators, but it influences the dynamics nonetheless. Indeed, all the eigenvalues of each subsystem play a role in the energies, but only the single eigenvalues $\nu_{1,2}$ are important for energy variations. These eigenvalues carry with them the collective effects of each subsystems, which arise particularly when the intra-couplings $g_{1,2}$ are non-negligible. This gives rise to two extreme cases:
\begin{itemize}
\item \textit{Non-collective regime --} This is the regime where the coupling between all the modes of a single subsystem is weak, 
namely $\nu_1\approx \epsilon_1$ and $\nu_2\approx \epsilon_2$. This is true whenever 
\begin{eqnarray}
    N_x g_x  \ll \omega_x \;  ,
\end{eqnarray}
which also means that $\nu_1\approx \omega_1$ and $\nu_2\approx \omega_2$. Therefore, the detuning $\Delta$ becomes the actual detuning of subsystem frequencies, and it is then the latter that determines the sign of interaction energy (along with the temperatures) and whether renormalized energies are greater or lesser than the bare energies.

\item \textit{Collective regime --} When the intra-couplings $g_{1,2}$ are non-negligible, collective effects due to the magnitude of $N_1$ and $N_2$ start to emerge. The extreme case is whenever the collective eigenvalue dominates over the others, $(\nu_x-\epsilon_x)\gg \epsilon_x$ . This is true whenever 
\begin{eqnarray}
    N_x g_x \gg \omega_x \;.
\end{eqnarray}
For large values of $N_x$, this corresponds to $\nu_x \approx N_x g_x$. 
Then, the effective detuning $\Delta$ is dictated by the relationship between these two quantities, and could be also different in sign with respect to the bare detuning $\omega_1-\omega_2$, drastically changing the behavior of all quantities. 
\end{itemize}

\section{Results}\label{sec:results}

In this section, we show the behavior of the different definitions of thermodynamic quantities across distinct regimes. We will both perform the comparison for the homogeneous coupling and frequencies case, whose fully analytical solution is presented in Sec.~\ref{sec:analytical}, and with a more realistic scenario, where the frequencies of each subsystem are distributed normally around a main value (given by $\omega_1$ and $\omega_2$, respectively), with standard deviation guided by the parameter $\sigma$ (such that the two standard deviations in the distributions are given by $\sigma_1 = \sigma \omega_1$ and $\sigma_2 = \sigma \omega_2$, respectively). This last case is treated via exact numerical diagonalization. 

Contrary to the case with identical frequencies, where the dynamics is purely oscillatory during the whole evolution, including a non-trivial distribution of the frequencies allows for a trend towards stationary and irreversible behavior -- seen, for example, through the appearance of plateaus in the change of internal energy. Of course, recurrences will still occur later in time due to the finite size of the systems. 

We will keep the regimes highlighted in the analytical case (Sec.~\ref{sec:regimes}) as a guideline also for the case of distributed frequencies. The first part of our analysis is focused on internal energy in all the approaches, as well as on the impact of the interaction energy. In the second part of our study, instead, we will focus on the behavior and interplay of work and heat contributions. For the entirety of this part, we will assume that the intra-coupling strength within both systems is the same $g_1=g_2=g$.

\subsection{Energy}
We first study the role of interaction energy, in particular showing that, in this paradigmatic case of two thermal sets of harmonic oscillators, it is not negligible even in the dispersive regime. We then discuss the structural differences between the asymmetric approaches to first law (interaction and bare approaches) and the minimal dissipation, focusing on the appearance of strong coupling peaks. We study the relevance of the (effective) detuning in determining the magnitude and relative difference between approaches, and discuss the impact of the eigenvalues distribution over relaxation-like effects.

\subsubsection{Role of interaction energy} \label{sec:res-int-role}
\begin{figure}[tp]
\centering
\includegraphics[width=\textwidth]{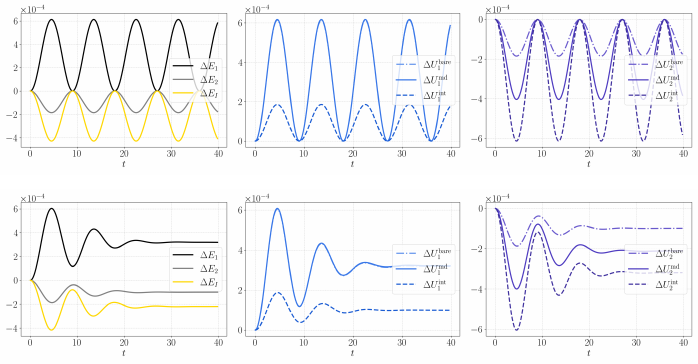}
\caption{Impact of interaction energy in the dispersive regime, homogeneous frequencies (upper row) and frequency distribution with $\sigma=0.1$ (lower row). (left): bare local energy variations and interaction energy variation, all of the same order of magnitude. (middle): internal-energy variation for subsystem 1 according to the three different definitions. In this case, the bare and minimal dissipation definitions are similar as a consequence of low system temperature. (right): internal-energy variation for subsystem 2 according to the three different definitions. In this case, all definition differ distinctly from each other. Parameters for all figures are $N=200$, $M=300$, $\omega_2 = 0.3 \omega_1$, $g=10^{-5} \omega_1$,$\gamma=10^{-5} \omega_1$, $T_1\omega_1 = 0.6$, $T_2\omega_1 = 4$. Horizontal axis shows time in units of $\omega_1^{-1}$.} 
\label{fig:plot-disp}
\end{figure}

We first notice that whenever $\nu_1=\nu_2$ the interaction energy is identically zero at all times for the analytical case of homogeneous frequencies. This condition is automatically satisfied when $\omega_1=\omega_2$, $N_1=N_2$, $g_1=g_2$, i.e. when the two subsystems are identical, except for their initial temperature. Then, the bare and interaction approach are identical by definition; in this case, however, the minimal dissipation approach is also identical to the others. Indeed, the magnitude of energy variations is dictated by $\nu_1 (n_2(\nu_1)-n_1(\nu_1))$, and therefore by the collective eigenvalue $\nu_1=\nu_2$ and the temperatures. The frequency of oscillation, instead, depends only on the coupling between the two subsystems (it is given by $N_1 \gamma = N_2 \gamma$). Moreover, when the two initial temperatures are also identical ($\beta_1=\beta_2$), all energy variations are identically zero. In this case, there is a possible build up of correlations between the two systems, but no dynamics at the level of the subsystems. Note, however, that these considerations are in general no longer valid when the frequencies of the subsystems are not homogeneous.

At weak coupling, the interaction energy between the two subsystems is expected to have less impact on the various definitions of thermodynamic quantities. To check this, we looked at the dispersive regime 
and found instead (in both the analytical case of fixed frequencies and the case of distributed frequencies) that the bare energy variations are of the same order of magnitude as the variation of interaction energy, see Fig.~\ref{fig:plot-disp}. As a direct consequence, the two sets of definitions that differ in how they account for the interaction energy (the ``bare'', and the ``interaction'' definitions) are incompatible with each other. 
Furthermore, one can see that also the minimal-dissipation definitions differ substantially from both the bare and the interaction definitions, as shown in  Fig.~\ref{fig:plot-disp} (particularly for subsystem 2).

From these analysis, we conclude that dispersive-like regimes (large detuning with respect to coupling) are not proper thermodynamic weak-coupling regimes, in the sense that
the interaction energy remains non-negligible, and this is evidenced by the fact that the various definitions adapted for the strong-coupling case 
predict distinct behaviors in this limit.

\subsubsection{Structural difference of the minimal dissipation approach}\label{sec:en-peaks}

As we have seen in the analytical case, the internal energies according to the bare and the interaction approaches (and, in general, the first law thermodynamic quantities) oscillate according to a single frequency $\Omega$ -- the analogue of the Rabi frequency. Because of the coefficient $B(t)$ appearing in the minimal dissipation internal energies, their behavior is much more complex and other secondary frequencies start to play a role, particularly at strong coupling. 

\begin{figure}[tp]
\centering 
\includegraphics[width=0.7\textwidth]{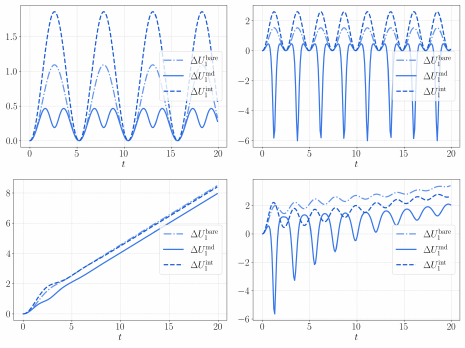} 
\caption{Secondary peaks in the minimal dissipation energy variation, for homogeneous frequencies (upper row) and large distribution of frequencies $\sigma=0.3$ (lower row). (left): internal-energy variation for subsystem 1 according to the three different definitions, at strong coupling $\gamma=2\cdot 10^{-3}\omega_1$. The minimal dissipation definition shows secondary dips which do not appear in the other definitions. (right):  internal-energy variation for subsystem 1 according to the three different definitions, at stronger coupling $\gamma=5 \cdot 10^{-3}\omega_1$. The dips in the minimal dissipation definition are drastically more pronounced.
Common parameters for the two figures are $N=200$, $M=300$, $\omega_2 = 1.7 \omega_1$, $g=10^{-5} \omega_1$, $T_1\omega_1 = 0.6$, $T_2\omega_1 = 4$. Horizontal axis shows time in units of $\omega_1^{-1}$.}
\label{fig:plot-dips}
\end{figure}

This effect mostly appears as formation of secondary peaks at half the main frequency. These peaks get stronger with stronger coupling, and are virtually unbounded the deeper the ultrastrong regime. We can see this analytically by looking at the ultrastrong coupling limit:
\begin{eqnarray}\label{eq:md-expansion-str}
\Delta U_1^{\mathrm{md}}(t) \stackrel{\Delta\ll \Gamma}{\approx} \frac{\nu}{2} [n_2(\nu_2)-n_1(\nu_1)]\sin^2\left(\frac{\Gamma}{2} t\right) + \frac{\Delta}{2} n_2(\nu_2) \tan^2\left(\frac{\Gamma}{2} t\right)
\end{eqnarray}
where the second term is divergent at times $\bar{t} = (2n+1)\pi/\Gamma$. This expansion becomes increasingly valid at stronger coupling, while for moderate coupling the peak smooths out earlier as mentioned in Sec.~\ref{sec:regimes}.
This feature is still present when the oscillator frequencies are spread around a main value. This can be seen in Fig.~\ref{fig:plot-dips}, where secondary peaks (dips, in this case) start to appear at medium couplings and become prominent at strong coupling.

\subsubsection{The relevance of detuning} 

\begin{figure}[tp]
\centering 
\includegraphics[width=0.7\textwidth]{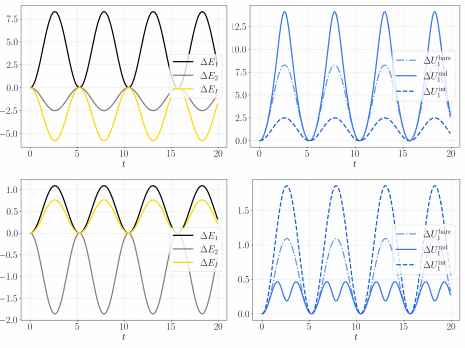}
\caption{Impact the detuning sign, homogeneous frequencies. (left column): bare local energy variations and interaction energy variation. (right column): internal-energy variation for subsystem 1 according to the three different definitions. (First row): positive detuning, taking $\omega_2 = 0.3\omega_1$.
(Second row): negative detuning, taking $\omega_2 = 1.7\omega_1$.
Interaction energy variation has the same sign as the detuning. The minimal dissipation definition is either larger or smaller than the bare definition depending on the detuning sign.
Common parameters for all figures are $N=200$, $M=300$, $g=10^{-5} \omega_1$,$\gamma=2\cdot 10^{-3} \omega_1$, $T_1\omega_1 = 0.6$, $T_2\omega_1 = 4$. Horizontal axis shows time in units of $\omega_1^{-1}$.}
\label{fig:plot-det-a}
\end{figure}

The effective detuning $\Delta$ is a crucial parameter for the thermodynamic quantities of the subsystems.
By looking at the terms of eq.~\eqref{eq:md-expansion-str} responsible for the secondary peaks, we also see that their sign depends on $\Delta$. 
Moreover, the detuning is also responsible, in the analytical case, for the sign of the interaction energy variation and the sign of difference between 
the bare-approach definition and the minimal-dissipation one (see Eqs.~\eqref{eq:Ubare-a} and \eqref{eq:Umd-a}). This can be seen explicitly in Fig.~\ref{fig:plot-det-a}, where we show two cases with opposite detuning (in the non-collective regime where the effective detuning is basically the bare detuning) but otherwise same parameters.

The case where the system frequencies are distributed around a value does not follow such a strict characterization, because of the multitude 
of actual frequency differences involved. Moreover, it is worth to point out that, contrary to what might be suggested by Fig.~\ref{fig:plot-det-a}, there is in principle no particular hierarchy between different definitions of internal energy. As an example, observe Fig.~\ref{fig:plot-dips}, lower right quadrant.

\begin{figure}[tp]
\centering 
\includegraphics[width=\textwidth]{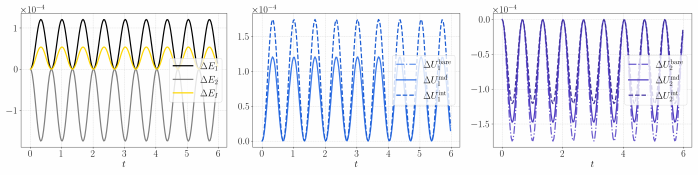}
\caption{Impact of interaction energy in the collective regime, homogeneous frequencies. (left): bare local energy variations and interaction energy variation, all of the same order of magnitude. (middle): internal-energy variation for subsystem 1 according to the three different definitions. (right): internal-energy variation for subsystem 2 according to the three different definitions. Parameters for all three figures are $N=200$, $M=300$, $\omega_2 = 0.3 \omega_1$, $g=10^{-1} \omega_1$,$\gamma=2\cdot 10^{-3} \omega_1$, $T_1\omega_1 = 0.6$, $T_2\omega_1 = 4$. Horizontal axis shows time in units of $\omega_1^{-1}$.}
\label{fig:plot-coll}
\end{figure}

Importantly, it is the effective detuning between the two subsystems -- i.e., the one given by the collective eigenvalues -- that dominates the effects described above. One way to see this is by moving to the collective regime (see Sec.~\ref{sec:regimes}), where the intra-coupling parameter $g$ (assumed the same for both baths) is strong enough. Depending on the parameter interplay, the effective detuning in this regime may have opposite sign with respect to the bare detuning $\omega_1-\omega_2$. Indeed, the bath with the largest number of modes has the largest effective eigenvalue, even if the bare frequency is the smaller of the two. In Fig.~\ref{fig:plot-coll}, we show a case of bare positive detuning and negative effective detuning. 

In the same figure, we observe that the modified effective detuning $\Delta$ leads to further drastic changes by comparing the plot with the top row of Fig.~\ref{fig:plot-det-a}. Since $\Delta$ features in the leading frequency $\Omega$, we witness a stark change of the oscillation frequency. As a consequence, this can lead to an effective change of regime -- in this instance, with respect to Fig.~\ref{fig:plot-det-a} we have effectively moved back into the dispersive regime (according to the definition of Sec.~\ref{sec:regimes}). 

\subsubsection{The impact of frequencies distribution}

As we mentioned before, the analytical scenario where all frequencies are the same cannot offer any sort of (even partially) irreversible behavior -- indeed, all quantities oscillate indefinitely. When distributing the frequencies around a main value, instead, one starts to see such effects, in a way that seems to be tied to the distribution of the eigenvalues of the total system.

\begin{figure}[tp]
\centering 
\includegraphics[width=\textwidth]{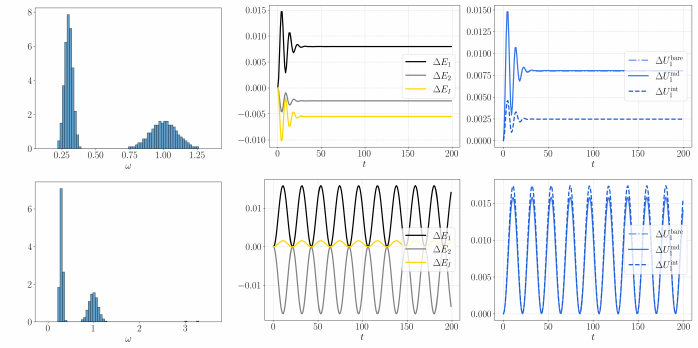}
\caption{Thermalization effects (or lack thereof) for distributed frequencies in the dispersive regime. Upper row shows non-collective regime ($g=10^{-5}\omega_1$) while lower shows collective ($g=10^{-2}\omega_1$). (left): Normalized histograms of eigenvalues (middle): bare local energy variations and interaction energy variation. (right): internal-energy variation for subsystem 1 according to the three different definitions. The non-collective regime shows the build-up of energy plateaus, while the collective regime shows persistent oscillations and an absence of plateaus. Parameters for all three figures are $N=200$, $M=300$, $\omega_2 = 0.3 \omega_1$,$\gamma=5\cdot 10^{-5} \omega_1$, $T_1\omega_1 = 0.6$, $T_2\omega_1 = 4$, $\sigma =0.1$. Horizontal axis shows time in units of $\omega_1^{-1}$.}
\label{fig:plot-thermal}
\end{figure}

We show thermalization signatures for a small increase in distribution in the dispersive regime. Fig.~\ref{fig:plot-thermal}, in particular, 
compares the case of small intra-coupling strength and the corresponding collective case (large intra-coupling parameter). The first regime shows signatures of thermalization (energies plateauing), while the collective case keeps oscillating with no visible creation of energy plateaus. In the figure, we show also the histograms of eigenvalues present in the entire system: the collective regime shows two eigenvalues (the leading, collective ones) appearing far-off from the main bulk of normal mode frequencies. These outlier modes contribute oscillatory components to the time evolution of the different quantities, with frequencies significantly different from those of the remaining modes, which leads to the lack of thermalization effects.
%

\subsection{Heat and work}

This part of our analysis instead focuses on work and heat contributions according to the three sets of definitions. We will focus first on the dispersive regime, where we find that the minimal dissipation approach is the only one compatible with the weak-coupling definition of heat, while the others show a significant deviation. This in turn shows that any difference between the internal energy of minimal dissipation and the one from the bare approach is, in the minimal dissipation definitions, assigned largely to work contributions. We further revisit the strong-coupling peaks found in the minimal dissipation approach, and see how these are mostly due to work-like contributions. Lastly, we show some cases of energy, work and heat balances for all definitions, 
showing how none of them satisfies a balance (not even in the dispersive regime), and that there is no ``ranking'' of definitions (meaning, a set of definitions that is more balanced than another).

\subsubsection{Heat in the dispersive regime}
Let us first recall that the heat contributions for the bare and interaction approaches are identical by definition. Moreover, in the case at hand where there is no explicit external driving on the systems, the definition of internal-energy variation of the bare approach coincides with the traditional definition of internal energy (and heat as well) employed in the weak-coupling approximations, 
\begin{equation}
     \Delta U_x^{\text{bare}}(t)=\Delta U_x^{\text{wc}}(t)=\delta Q_x^{\text{wc}}(t) =  E_x(t)-E_x(0).
\end{equation}

We already showed in Sec.~\ref{sec:res-int-role} that in the dispersive regime the interaction energy variation is still non-negligible with respect to the bare energy variations. Therefore, the bare/interaction definition of heat (which accounts for interaction-energy variation) is for sure quantitatively different from the heat contribution in the weak-coupling definitions.
However, this is not true for the minimal dissipation heat, which is instead compatible with the weak-coupling definition. This can be seen by performing the relevant expansion of heat in the dispersive regime given the analytical formulas:
\begin{eqnarray}
    \delta Q_x^{\text{md}}(t)  \stackrel{\Delta \gg \Gamma}{\approx} \nu_x [ n_{\bar{x}}(\nu_{\bar{x}})- n_{{x}}(\nu_{{x}})]\frac{\Gamma^2}{\Delta^2}\sin^2\left( \frac{\Delta t}{2}\right)  \stackrel{\Delta \gg \Gamma}{\approx} \delta Q_x^{\text{wc}}(t)\; .
\end{eqnarray}

\begin{figure}[tp]
\centering
\includegraphics[width=0.7\textwidth]{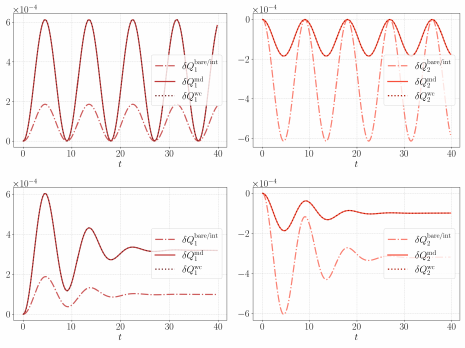}
\caption{Heat contributions in the dispersive regime, homogeneous frequencies (upper row) and slight frequency distribution $\sigma=0.1$ (lower row). (left): heat exchange for subsystem 1 according to the three different definitions (bare and interaction coincide by definition) and the weak-coupling definition. (right): heat exchange for subsystem 2 according to the different definitions. Curves from the weak-coupling approximation and the minimal dissipation approach overlap. Parameters for all figures are $N=200$, $M=300$, $\omega_2 = 0.3 \omega_1$, $g=10^{-5} \omega_1$,$\gamma=10^{-5} \omega_1$, $T_1\omega_1 = 0.6$, $T_2\omega_1 = 4$. Horizontal axis shows time in units of $\omega_1^{-1}$.}
\label{fig:plot-disp-heat}
\end{figure}

\begin{figure}[tp]
\centering
\includegraphics[width=0.7\textwidth]{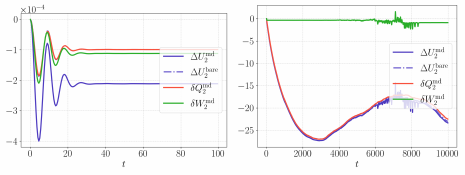}
\caption{Comparison between bare internal energy and minimal dissipation contributions in the dispersive regime. (left): small frequency distribution ($\sigma=0.1$), couplings $g=10^{-5} \omega_1$ and $\gamma=10^{-5} \omega_1$. (right): larger frequency distribution ($\sigma=0.3$), couplings $g=10^{-4} \omega_1$ and $\gamma=5\cdot10^{-4} \omega_1$ Parameters for all figures are $N=200$, $M=300$, $\omega_2 = 0.3 \omega_1$, $T_1\omega_1 = 0.6$, $T_2\omega_1 = 4$. Horizontal axis shows time in units of $\omega_1^{-1}$. When not visible, $\Delta U_2^{\text{bare}}$ overlaps with $\delta Q_2^{\text{md}}$.}
\label{fig:plot-disp-comparison}
\end{figure}

We can see this feature reflected in the plots also beyond the analytical case, see Fig.~\ref{fig:plot-disp-heat}.
Therefore, this also leads to the observation that the difference between the minimal dissipation internal energy and the internal energy given by the bare approach (equal to the weak-coupling heat) is given just by work contributions; see also Fig.~\ref{fig:plot-disp-comparison}.

\subsubsection{Work peaks in the minimal dissipation approach}

In Sec.~\ref{sec:en-peaks}, we highlighted an extra structure appearing in the minimal dissipation definition of internal energy, namely additional peaks (or dips, depending on detuning sign) that get larger in magnitude the more one moves towards the ultrastrong coupling regime. 
A natural question to ask, from the perspective of the first law, is whether these constitute mostly heat or work contributions. From the analytical calculations and the strong coupling expansion of work and heat, it is possible to infer that these are mainly due to work. The expansions for the first subsystem read:
\begin{eqnarray}
    \hspace{-2em}\delta Q_1^{\text{md}}(t)  \stackrel{\Delta \ll \Gamma}{\approx} [n_2(\nu_2)-n_1(\nu_1)]\left[\frac{\nu}{2}\sin^2\left(\frac{\Gamma}{2} t\right) - \ln\left(\left| \cos\left( \frac{\Gamma t}{2}\right)\right| \right) \right], &&\\
    \hspace{-2em}\delta W_1^{\text{md}}(t)  \stackrel{\Delta \ll \Gamma}{\approx} [n_2(\nu_2)-n_1(\nu_1)]\ln\left(\left| \cos\left( \frac{\Gamma t}{2}\right)\right| \right) + \frac{\Delta}{2} n_2(\nu_2) \tan^2\left(\frac{\Gamma}{2} t\right) .&&
\end{eqnarray}
As one can see from above, both contributions feature divergences at $\bar{t} = (2n+1)\pi/\Gamma$, but the terms containing the logarithm (which cancel out in the internal-energy variation) are slower, and can be expected to be smoothed out much faster than the divergence contained in the tangent term -- which is responsible for the peaks in the internal energy (see Eq.~\ref{eq:md-expansion-str}). Indeed, this term is only present in the work contribution. 

\begin{figure}[tp]
\centering 
\includegraphics[width=0.7\textwidth]{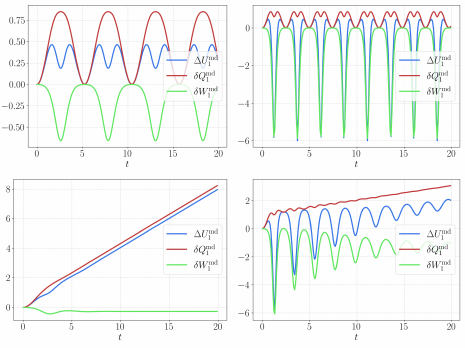} 
\caption{Work peaks in the minimal dissipation energy variation, for homogeneous frequencies (upper row) and large distribution of frequencies $\sigma=0.3$ (lower row). (left): minimal dissipation quantities for subsystem 1, at strong coupling $\gamma=2\cdot 10^{-3}\omega_1$. (right):  minimal dissipation quantities for subsystem 1, at stronger coupling $\gamma=5 \cdot 10^{-3}\omega_1$. The dips in internal energy in all cases are mostly due to work-like contributions.
Common parameters for the two figures are $N=200$, $M=300$, $\omega_2 = 1.7 \omega_1$, $g=10^{-5} \omega_1$, $T_1\omega_1 = 0.6$, $T_2\omega_1 = 4$. Horizontal axis shows time in units of $\omega_1^{-1}$.}
\label{fig:plot-dips-work}
\end{figure}

Looking at the plots, the intuition given by the expansion in confirmed, and the feature turns out to be stable also for the non-analytical case of distributed frequencies, see Fig.~\ref{fig:plot-dips-work}.

\subsubsection{Energy balances}
From classical macroscopic thermodynamics, one is used to seeing conserved flows of energy between two sides of a bipartite system. Namely, one expects that if one side of the system loses a certain amount of internal energy (or heat, or work), the other side of the system gains the same amount of internal energy (heat, work). This is ensured by the fact that the interaction energy is negligible in typical macroscopic limits. In quantum thermodynamics, especially at the strong-coupling regime, the interaction energy is no longer negligible and thus such intuitive relations should not be expected to hold.

This is indeed the case for our study. Defining the net energy exchanges
\begin{eqnarray}
\overline{\Delta U}(t) &=& \Delta U_1(t) +  \Delta U_2(t) \\
\overline{\delta Q}(t) &=& \delta Q_1(t) +  \delta Q_2(t) \\
\overline{\delta W}(t) &=& \delta W_1(t) +  \delta W_2(t) 
\end{eqnarray}
for each set of definitions considered in this work, we know that none of these quantities can be expected to be equal to zero (particularly for the bare and interaction approaches, as well as the weak-coupling definitions, but also true for the minimal dissipation definitions). 
Nonetheless, energy additivity -- in the sense of satisfying balance equations -- is at times regarded as a requirement for a satisfactory theory of thermodynamics at strong coupling \cite{Neves2024}. We here briefly investigate whether such balances hold at least approximately, in the case of the two ``baths''. 

\begin{figure}[tp]
\centering 
\includegraphics[width=\textwidth]{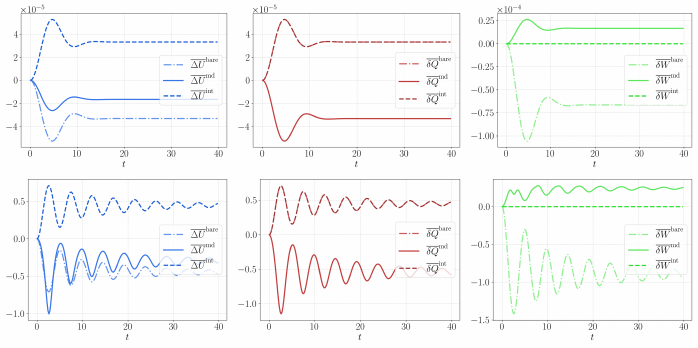} 
\caption{Energy (left), heat (middle) and work (left) balances for distributed frequencies ($\sigma=0.1$) for all three definitions. Rows show coupling strength $\gamma=10^{-5}$, $\gamma=2\cdot10^{-3}$.
Common parameters for the two figures are $N=200$, $M=300$, $\omega_2 = 1.7 \omega_1$, $g=10^{-5} \omega_1$, $T_1\omega_1 = 0.6$, $T_2\omega_1 = 4$. Horizontal axis shows time in units of $\omega_1^{-1}$.}
\label{fig:plot-balance-2}
\end{figure}

We plot internal energy balance, heat balance, and work balance for all three sets of approaches in Fig.~\ref{fig:plot-balance-2}. We can see that all of the balances are indeed different from zero, both in the dispersive regime and in the strong coupling regime (except for the interaction-approach work contribution, which is identically zero for both systems by definition). While the order of magnitude of the balances seems to be much smaller in the dispersive regime, we remark that it is fully compatible with the order of magnitude of the actual energy exchanges for those parameter settings. Indeed, the non-zero balances of the various quantities and approaches are all of the order of $20\%-50\%$ of the respective energy variations. A notable exception is the work balance for the bare approach, which includes twice the interaction-energy variation (recall Eq.~\eqref{eq:balance-bare}).

\section{Conclusions}\label{sec:con}

We have comprehensively compared three different sets of definitions for internal energy, heat, and work in a model of two interacting thermal subsystems, where there is no natural system-bath distinction. Our analysis shows that these definitions -- two asymmetric ones commonly used in the literature, and a third based on the minimal dissipation framework -- lead to substantially different predictions, even in regimes of weak coupling and large detuning. This emphasizes that dispersive conditions alone are not sufficient to recover a universal notion of energy exchange, that interaction energy contributions remain relevant, and that different definitions encode them in incompatible ways.

Among the three approaches, the minimal dissipation framework shows additional dynamical features (secondary peaks) in the internal energy, that are absent in the others and grow dominant in the ultrastrong coupling regime. We traced these structures back to work-like contributions. Whether these work-like contributions can be harnessed for concrete tasks -- such as energy harvesting, information processing, or control at the quantum level -- remains an open question. Investigating their operational relevance and the conditions under which they can be systematically exploited represents a key avenue for future research. Moreover, we found that only the minimal dissipation definition of heat reduces to the standard weak-coupling form in the dispersive limit, suggesting that it consistently extends weak-coupling thermodynamics.

We also observed that the sign and magnitude of energy flows are controlled not by bare detuning, but by effective detuning derived from the collective spectrum of the full system. This leads to unexpected reversals and qualitative shifts in behavior, particularly in the collective regime. Finally, none of the definitions examined satisfy an energy balance across the bipartition -- not even approximately -- highlighting the conceptual and practical ambiguities that arise when interaction energies cannot be neglected.

\section*{Acknowledgements}

This work has been supported by the Italian Ministry of Research and Next Generation EU via the NQSTI-Spoke1-BaC project QSynKrono (contract n. PE00000023-QuSynKrono) and via the PRIN 2022 project Quantum Reservoir Computing (QuReCo) (contract
n. 2022FEXLYB).

\section*{References}
\bibliography{biblio}

\begin{appendix}
\section{Analytical derivation for homogeneous coupling and frequencies}\label{sec:appA}
Here we show the derivation of the quantities reported in Sec.~\ref{sec:analytical}. Let us first set some notation by defining the matrix
\begin{eqnarray}
    J_{N_1 N_2} =
\left[\begin{array}{cccc}
1 & 1 & \cdots & 1 \\
1 & 1 & \cdots & 1 \\
\vdots & \vdots & \ddots & \vdots \\
1 & 1 & \cdots & 1
\end{array}\right]_{N_1 \times N_2}
\end{eqnarray}
with short-hand notation $J_N := J_{NN}$.
The full Hamiltonian matrix $H$ is then given by the following blocks
\begin{eqnarray}
    H =
\left[\begin{array}{cc}
H_1 &  \gamma J_{N_1 N_2} \\
\gamma J_{N_2 N_1} & H_2
\end{array} \right] \; ,
\end{eqnarray}
where the bare Hamiltonian matrices of the two subsets are given by
\begin{eqnarray}
    H_1 &= (\omega_1-g_1)\mathbb{I}_{N_1} + g_1 J_{N_1} \; , \\
    H_2 &= (\omega_2-g_2)\mathbb{I}_{N_2} + g_2 J_{N_2} \; .
\end{eqnarray}

\textbf{Eigensystem of} $\boldsymbol{H_x}$ --- 
We first diagonalize the blocks related to the two bare Hamiltonians. There are $N_x-1$ degenerate eigenvectors $\vec{v}_1, ... \vec{v}_{N_x-1}$, which we arrange to form an $N_x\times(N_x-1)$ matrix 
\begin{eqnarray}
    V_{N_x} := \big[ \vec{v}_1, ...,  \vec{v}_{N_x-1} \big]
\end{eqnarray}
of eigenvalue
\begin{eqnarray}
    \epsilon_x = \omega_x - g_x.
\end{eqnarray}
Then, there is one eigenvector $\vec{v}_{N_x} = \frac{1}{\sqrt{N_x}}[1,1,...,1]^T$ of eigenvalue
\begin{eqnarray}
    \nu_x = \omega_x + (N_x-1)g_x
\end{eqnarray}
Then, the matrix diagonalizing $H_x$ is given by $P_x= [V_{N_x}, \vec{v}_{N_x}]$. It's useful to notice, for later, that the two following identities hold:
\begin{eqnarray}\label{eq:an-rel-v1}
    \vec{v}_{N_x} \vec{v}_{N_x}^T &= \frac{1}{N_x} J_{N_x} \; , \\ \label{eq:an-rel-v2}
    V_{N_x} V_{N_x}^T &= \mathbb{I}_{N_x} - \vec{v}_{N_x} \vec{v}_{N_x}^T = \mathbb{I}_{N_x} - \frac{1}{N_x} J_{N_x} \; .
\end{eqnarray}

\textbf{Eigensystem of }$\boldsymbol{H}$ --- 
We now diagonalize the global system Hamiltonian, including the interaction.
There are $N_1-1$ degenerate eigenvectors -- the same as for $H_1$, namely the ones arranged in $V_{N_1}$ (but extended with zeroes to become $N_1+N_2$ dimensional vectors) -- of eigenvalue $\epsilon_1$. The $(N_1+N_2)\times (N_1-1)$ matrix of these eigenvectors is 
\begin{eqnarray}
    \left[\begin{array}{c}
            V_{N_1} \\
            \mathbb{0}_{N_2\times(N_1-1)}
    \end{array}\right] \; .
\end{eqnarray}
Similarly, there are $N_2-1$ degenerate eigenvectors -- the same as for $H_2$ -- of eigenvalue $\epsilon_2$. The $(N_1+N_2)\times (N_2-1)$ matrix of these eigenvectors is 
\begin{eqnarray}
    \left[\begin{array}{c}
            \mathbb{0}_{N_1\times(N_2-1)} \\
            V_{N_2}
        \end{array}\right] \; .
\end{eqnarray}
Then, there is one eigenvector of eigenvalue
\begin{eqnarray}
    \lambda= \frac{\nu +\Omega}{2} \; ,
\end{eqnarray}
where we have defined $\nu$ and $\Omega$ as in Eq.(\ref{eq:nuomega}).
The eigenvector associated to $\lambda$ is the following
    \begin{eqnarray}
        \left[\begin{array}{c}
            f_+ \vec{v}_{N_1} \\
             f_- \vec{v}_{N_2}
        \end{array}\right] \; ,
    \end{eqnarray}
with coefficients
    \begin{eqnarray}
        f_\pm = \frac{1}{\sqrt{2}} \sqrt{1 \pm \frac{\Delta}{\Omega}}\; .
    \end{eqnarray}
Lastly, there is one eigenvector of eigenvalue
    \begin{eqnarray}
        \mu = \frac{\nu -\Omega}{2} \; ,
    \end{eqnarray}
which reads
    \begin{eqnarray}
        \left[\begin{array}{c}
            -f_- \vec{v}_{N_1} \\
             f_+ \vec{v}_{N_2}
        \end{array} \right] \; .
    \end{eqnarray}
Therefore, the matrix $Z$ that diagonalizes $H$ (with convention $H'= Z^T HZ$) is given by
\begin{eqnarray}\label{eq:an-Z}
    Z = \left[\begin{array}{cccc}
        V_{N_1} & f_+ \vec{v}_{N_1} & \mathbb{0} & -f_- \vec{v}_{N_1} \\
        \mathbb{0} & f_- \vec{v}_{N_2} & V_{N_2} & f_+ \vec{v}_{N_2}
    \end{array}\right] \; ,
\end{eqnarray}
while the diagonalized matrix reads
\begin{eqnarray}\label{eq:an-H'}
    H' = \left[\begin{array}{cccc}
        \epsilon_1 \mathbb{I}_{N_1-1} & \mathbb{0} & \mathbb{0}&\mathbb{0} \\
       \mathbb{0} & \lambda & \mathbb{0} & \mathbb{0}  \\
        \mathbb{0} & \mathbb{0}&  \epsilon_2 \mathbb{I}_{N_2-1} &\mathbb{0} \\   
\mathbb{0} & \mathbb{0} & \mathbb{0} &\mu
    \end{array}\right] \; .
\end{eqnarray}

\textbf{Total unitary transformation }$\boldsymbol{U_t}$ --- 
We can now use the global eigensystem and the diagonalizing matrix $Z$ to obtain the matrix representing the unitary evolution of the system, namely $U_t = Z e^{-iH't}Z^T$. By using the block expressions \eqref{eq:an-Z}, \eqref{eq:an-H'}, and keeping in mind the relations \eqref{eq:an-rel-v1} and \eqref{eq:an-rel-v2}, we write it in block form
    \begin{eqnarray}
        U_t = \left[\begin{array}{cc}
            U_{11} & U_{12} \\
            U_{21} & U_{22}
        \end{array}\right]
    \end{eqnarray}
where we have
    \begin{eqnarray}
        U_{11} &= e^{-i \epsilon_1 t}\mathbb{I}_{N_1} + \frac{1}{N_1}(\alpha_1(t) - e^{-i \epsilon_1 t}) J_{N_1}\\
        U_{12} &= \frac{1}{\sqrt{N_1 N_2}}\xi(t) J_{N_1 N_2}\\
        U_{21} &= \frac{1}{\sqrt{N_1 N_2}}\xi(t) J_{N_2 N_1}\\
        U_{22} &= e^{-i \epsilon_2 t}\mathbb{I}_{N_2} + \frac{1}{N_2}(\alpha_2(t) - e^{-i \epsilon_2 t}) J_{N_2}
    \end{eqnarray}
and with the newly defined quantities
    \begin{eqnarray}
        \alpha_1(t) &= e^{-i\lambda t}f_+^2 +e^{-i\mu t}f_-^2 \\
        \alpha_2(t) &= e^{-i\lambda t}f_-^2 +e^{-i\mu t}f_+^2  \\
        \xi(t) &= f_+ f_- (e^{-i\lambda t}- e^{-i\mu t}) \; .
    \end{eqnarray}

\textbf{Matrix of $\boldsymbol{\braket{a^\dag_ia_j}_0}$ for an initial thermal state} ---
To find the energies, we need to know the matrix $S_t=\braket{\vec{R}^\dag(\vec{R})^T}_t$. Before we evolve it, we need to find its initial condition $S_0 = \braket{\vec{R}^\dag(\vec{R})^T}_0$ for which the two subsystems are initially in a thermal state (but uncoupled). We do this again by finding the normal modes, this time imposing $\gamma=0$. Then, the Hamiltonian matrix is diagonalized by 
    \begin{eqnarray}
        \tilde{Z} = \left[\begin{array}{cc}
            P_1 & \mathbb{0}\\
            \mathbb{0} & P_2 
        \end{array}\right] \; ,
    \end{eqnarray}
which transforms the subsystems into normal modes -- new coordinates $R'$ for which
    \begin{eqnarray}
    S'_0 =\braket{\vec{R'}^\dag(\vec{R'})^T}_0= \left[\begin{array}{cc}
            \mathcal{N}'_1 & \mathbb{0}\\
            \mathbb{0} & \mathcal{N}'_2
        \end{array}\right] \; ,
    \end{eqnarray}    
where the matrices $\mathcal{N}'_{1,2}$ are diagonal and encode the average excitation numbers of the normal modes according to a thermal distribution with inverse temperature $\beta_{1,2}$; namely,
\begin{eqnarray}
    \mathcal{N}'_x = \left[\begin{array}{cc}
            n_x(\epsilon_x)\mathbb{I}_{N_x-1} & \mathbb{0}\\
            \mathbb{0} & n_x(\nu_x)
        \end{array} \right]\; ,
\end{eqnarray}   
where $n_x(X)= (e^{\beta_x X}-1)^{-1}$.
Using $\vec{R}' = \tilde{Z}^T \vec{R}$, we can write
    \begin{eqnarray}
    S_0= \tilde{Z} S_0' \tilde{Z}^T = \left[\begin{array}{cc}
            \mathcal{N}_1 & \mathbb{0}\\
            \mathbb{0} & \mathcal{N}_2
        \end{array}\right] \; ,
    \end{eqnarray}   
where
    \begin{eqnarray}
        \mathcal{N}_x &= n_x(\epsilon_x) \mathbb{I}_{N_x} + \frac{1}{N_x}\left( n_x(\nu_x) -  n_x(\epsilon_x)\right)J_{N_x} \;  . 
    \end{eqnarray}

\textbf{Matrix of $\boldsymbol{\braket{a^\dag_ia_j}_t}$ evolved in time} ---
We can now evolve $S_0$ using equation \eqref{eq:S-evolved}. It gives
    \begin{eqnarray}
      \hspace*{-2em} S_t = U_t^* S_0 U_t = \left[\begin{array}{cc}
            (U^*_{11}N_1U_{11}+ U^*_{12}N_2U_{21}) & (U^*_{11}N_1U_{12}+U^*_{12}N_2U_{22}) \\
            (U^*_{21}N_1U_{11}+U^*_{22}N_2U_{21})& (U^*_{21}N_1U_{12}+ U^*_{22}N_2U_{22})
        \end{array}\right] \; ,
    \end{eqnarray}
showing that correlations have built up between the two subsystems. For the sake of calculating the energies, though, we only need the reduced matrix for each subsystem. These are given by
    \begin{eqnarray}
        S_t^{(1)} &= U^*_{11}N_1U_{11}+ U^*_{12}N_2U_{21} \; , \\
        S_t^{(2)} &=U^*_{21}N_1U_{12}+ U^*_{22}N_2U_{22} \; .
    \end{eqnarray}
Notice that the following relation holds for a product of two matrices of the form $X= A\mathbb{I}_N + (B-A)J_N/N$:
    \begin{eqnarray}\label{eq:an-multiplication}
        &&(A\mathbb{I}_N + \frac{1}{N}(B-A)J_N)\cdot(C\mathbb{I}_N + \frac{1}{N}(D-C)J_N) \nonumber\\
        &&= AC\mathbb{I}_N + \frac{1}{N}(BD-AC)J_N \; .
    \end{eqnarray}
Using the above and the expressions for the blocks of $U_t$ we find 
    \begin{eqnarray}
      \hspace*{-1em}  S_t^{(x)} = n_x(\epsilon_x) \mathbb{I}_{N_x} + \frac{1}{N_x}\left( n_x(\nu_x)|\alpha(t)|^2 + n_{\bar{x}}(\nu_{\bar{x}})|\xi(t)|^2 -  n_x(\epsilon_x)\right)J_{N_x}   ,
    \end{eqnarray}
where we have introduced  
\begin{eqnarray}
|\alpha(t)|^2:=|\alpha_1(t)|^2=|\alpha_2(t)|^2= 1-|\xi(t)|^2 \; .
\end{eqnarray}

\textbf{Renormalized Hamiltonians }$\boldsymbol{K_t}$ ---
We are now ready to compute the renormalized Hamiltonians for both subsystems using formulas \eqref{eq:Mt-firts} and \eqref{eq:Kt} and the unitary evolution just found. We will briefly show the steps to obtain the Hamiltonian $K_t$ for subsystem 1. The one for the second subsystem can be obtained analogously.
The matrix $L_t$ reads
    \begin{eqnarray}
        L_t = \dot{U}_{11}U_{11}^{-1} \; .
    \end{eqnarray}
To evaluate the time-derivative and the inverse of $U_{11}$, we notice that given any matrix of the form $X= A\mathbb{I}_N + (B-A)J_N/N$ it holds
    \begin{eqnarray}
        \dot{X} &= \dot{A}\mathbb{I}_N + \frac{1}{N}\left(\dot{B}-\dot{A}\right)J_N \\
        X^{-1} &= \frac{1}{A}\mathbb{I}_N + \frac{1}{N}\left(\frac{1}{B}-\frac{1}{A}\right)J_N \; .
    \end{eqnarray}
For the matrix $U_{11}$, the structure is the one above, with $A= e^{-i\epsilon_1 t}$ and $B= \alpha_1(t)$. Using also Eq.~(\ref{eq:an-multiplication}), we obtain 
    \begin{eqnarray}
        L_t = - i \epsilon_1 \mathbb{I}_{N_1} + \frac{1}{N_1}\left(\frac{\dot{\alpha}_1(t)}{\alpha_1(t)}+i \epsilon_1\right)J_{N_1} \; ,
    \end{eqnarray}
which gives the renormalized Hamiltonian 
    \begin{eqnarray}
        K_t^{(1)} = \epsilon_1 \mathbb{I}_{N_1} + \frac{1}{N_1}\left(-\Im\left\{\frac{\dot{\alpha}_1(t)}{\alpha_1(t)}\right\}- \epsilon_1\right)J_{N_1} \; .
    \end{eqnarray}
By analogy, the renormalized Hamiltonian for the second subsystem reads
    \begin{eqnarray}
        K_t^{(2)} = \epsilon_2 \mathbb{I}_{N_2} + \frac{1}{N_2}\left(-\Im\left\{\frac{\dot{\alpha}_2(t)}{\alpha_2(t)}\right\}- \epsilon_2\right)J_{N_2} \; ,
    \end{eqnarray}
where we find
    \begin{eqnarray}
        \Im\left\{\frac{\dot{\alpha}_1(t)}{\alpha_1(t)}\right\} &= -\frac{\nu}{2} - \frac{\Delta}{2|\alpha(t)|^2} \; , \\
        \Im\left\{\frac{\dot{\alpha}_2(t)}{\alpha_2(t)}\right\} &= -\frac{\nu}{2} + \frac{\Delta}{2|\alpha(t)|^2} \; .
    \end{eqnarray}

\textbf{Subsystem energies} --- 
All energies of interest are given in terms of the quantities
\begin{eqnarray}
    E_x(t) &= \Tr\{H_x S_t^{(x)}\} \; , \\
    U_x(t) &= \Tr\{K_t^{(x)} S_t^{(x)}\}\; .
\end{eqnarray}
They can be easily found using the multiplication formula \eqref{eq:an-multiplication} and the fact that $\Tr\{\mathbb{I}_N\}=\Tr\{J_N\}=N$. Massaging the expressions we eventually find
\begin{eqnarray}
   \hspace{-0.15cm} E_1(t) &= (N_1-1)\epsilon_1 n_1(\epsilon_1) +\nu_1 [n_1(\nu_1)+G(t)] , \\
   \hspace{-0.15cm} E_2(t) &= (N_2-1)\epsilon_2 n_2(\epsilon_2) +\nu_2 [n_2(\nu_2)-G(t)] , \\
   \hspace{-0.15cm} U_1(t) &= (N_1-1)\epsilon_1 n_1(\epsilon_1) + [\nu_1(1-B(t)) + \nu_2 B(t)]\cdot[n_1(\nu_1)+G(t)] , \\
   \hspace{-0.15cm} U_2(t) &= (N_2-1)\epsilon_2 n_2(\epsilon_2) + [\nu_1 B(t) + \nu_2 (1-B(t))]\cdot [n_2(\nu_2)-G(t)] . 
\end{eqnarray}
In the above, we have defined the following quantities
\begin{eqnarray}
    G(t) &= (n_2(\nu_2) - n_1(\nu_1))|\xi(t)|^2 \; , \\
    |\xi(t)|^2 &= \frac{\Gamma^2}{\Omega^2} \sin^2\left(\frac{\Omega}{2} t\right)\; , \\
    B(t) &= -\frac{\Gamma^2 \sin^2\left(\frac{\Omega}{2} t\right)}{2\left[\Delta^2+\Gamma^2\cos^2\left(\frac{\Omega}{2} t\right)\right]}  \; . 
\end{eqnarray}
By noticing that at initial times $G(0)=B(0)=0$, we have
\begin{eqnarray}
    E_x(0) = U_x(0)=(N_x-1)\epsilon_x n_x(\epsilon_x) +\nu_x n_x(\nu_x) \;  . 
\end{eqnarray}
This allows us to compute energy variations
\begin{eqnarray}
    \Delta E_1(t) &= +\nu_1 G(t) \; , \\
    \Delta E_2(t) &= -\nu_2 G(t) \; , \\
    \Delta U_1(t) &= +\nu_1 G(t) -B(t)\Delta[n_1(\nu_1)+G(t)]  \; , \\
    \Delta U_2(t) &= -\nu_2 G(t) + B(t)\Delta [n_2(\nu_2)-G(t)] \; ,
\end{eqnarray}
which in turn can be used to find the interaction energy variations according to the three different sets of definitions, as given in Sec.~\ref{sec:analytical}.

\end{appendix}

\end{document}